\documentclass[prb,preprint,preprintnumbers,amsmath,amssymb]{revtex4}

\usepackage{graphicx}

\usepackage{dcolumn}

\usepackage{bm}

\begin{document}


\title{Internal magnetic fields in thin ZnSe epilayers}

\author{S. Ghosh}
\author{N. P. Stern}
\author{B. Maertz}
\author{D. D. Awschalom}
\email{awsch@physics.ucsb.edu}

\affiliation{Center for Spintronics and Quantum Computation,
University of California, Santa Barbara, CA 93106 USA}

\author{G. Xiang}
\author{M. Zhu}
\author{N. Samarth}
\affiliation{Department of Physics, Pennsylvania State University,
University Park, Pennsylvania 16802 USA}

\date{\today}

\begin{abstract}

Strain induced spin-splitting is observed and characterized using
pump-probe Kerr rotation spectroscopy in $\it{n}$-ZnSe epilayers
grown on GaAs substrates. The spin-splitting energies are mapped
out as a function of pump-probe separation, applied voltage, and
temperature in a series of samples of varying epilayer thicknesses
and compressive strain arising from epilayer-substrate lattice
mismatch. The strain is independently quantified using
photoluminescence and x-ray diffraction measurements. We observe
that the magnitude of the spin-splitting increases with applied
voltage and temperature, and is highly crystal direction
dependent, vanishing along [1$\bar{1}$0].
\end{abstract}

\maketitle

ZnSe has been extensively investigated in the last decade for
applications in optoelectronic devices operational in the blue
spectral region$^{\cite{hasse}}$. More recently, the potential for
use of ZnSe in spintronics devices has been evidenced by the long
spin coherence times ($\sim$100 ns) in $\it{n}$-ZnSe and the
ability to coherently transfer spins across a heterointerface
between $\it{n}$-ZnSe and GaAs.$^{\cite{malajovich1},\cite{
malajovich2}}$ Experiments have further demonstrated electrical
generation of spins in $\it{n}$-ZnSe through both current-induced
spin polarization and the spin Hall Effect$^{\cite{stern}}$ at
room temperature. \par

Earlier studies in strained GaAs and InGaAs$^{\cite{kato1},\cite{
sih}}$ have revealed a $\it{k}$-linear spin-splitting which acts on
spins as an internally generated magnetic field . Despite the
observation of electrically generated spins, no internal magnetic
fields have been measured in ZnSe epilayers of micron-scale
thickness. In this letter, we investigate the presence of
electrically-induced internal magnetic field, \textit{B}$_{int}$, in
strained \textit{n}-type ZnSe epilayers. We observe an effective
\textit{B}$_{int}$ proportional to the drift velocity of electrons
and note that, although the bulk inversion asymmetry spin-orbit
parameter in ZnSe is almost half that in GaAs,$^{\cite{winkler}}$
the spin-splitting energy scale is comparable. Coupled with the
previous studies, this establishes the potential for developing
all-electrical protocols for spin initialization and coherent
manipulation in this wide band gap material.    \par

We measure three Cl-doped $\it{n}$-type ZnSe epilayers grown by
molecular beam epitaxy on semi-insulating (001) GaAs substrates.
The epilayer thicknesses are 100 nm, 150 nm and 300 nm with
carrier densities in the range of 4-5 $\times$ 10$^{17}$cm$^{-3}$.
Measurements of similar samples with 1.5-$\mu$m thick ZnSe
epilayers and similar carrier concentrations show no evidence of
an internal magnetic field$^{\cite{stern}}$. Figure 1a is an
optical image of a patterned sample, showing channels etched along
[110] and [1$\bar{1}$0] directions. The channels have width
$\it{w}$ = 100 $\mu$m, length $\it{l}$ = 235 $\mu$m along [110]
(along [1$\bar{1}$0], $\it{l}$ = 350 $\mu$m), and thickness
$\it{h}$ = 100, 150 and 300 nm. Annealed indium contact pads at
the ends of each arm allow us to independently apply an electric
field \textit{E} along both orthogonal directions.  \par

Growth of ZnSe on GaAs causes in-plane compressive strain in the
ZnSe film due to the small lattice mismatch between the two
zinc-blende semiconductors (0.25$\%$ at 300 K)$^{\cite{madelung}}$.
X-ray diffraction (XRD) measurements have shown that although
biaxial strain persists in epilayers as thick as 1.6
$\mu$m$^{\cite{kontos}}$, the critical thickness beyond which
dislocations begin to form in the ZnSe/GaAs system is about 0.17
$\mu$m$^{\cite{ohkawa}}$. We quantify the strain in our samples
using both photoluminescence (PL) and XRD measurements. Fig.1b shows
PL at $\it{T}$ = 50 K for the 300-nm and 100-nm samples. Comparing
the direct band gap of a 2-$\mu$m ZnSe epilayer with similar carrier
concentration at 50 K$^{\cite{wang}}$ (dashed line in the figure)
with the observed PL, we notice that the PL peaks are increasingly
red-shifted relative to the unstrained band gap with decreasing
epilayer thickness. By assuming that the PL peak occurs at the
direct bang gap energy, we can use the shift in the PL to estimate
the in-plane strain in the samples$^{\cite{mohammed}}$. To first
order, these calculations predict compressive strain in the range of
1$\times$10$^{-3}$ to 3$\times$10$^{-3}$, which are comparable to
the values quoted in existing measurements of samples of similar
thicknesses$^{\cite{kontos}}$. XRD measurements (fig. 1c and 1d)
show further evidence of monotonically increasing strain with
decreasing epilayer thickness, again in agreement with previous
studies. We do not observe any anisotropy in the strain about [115]
and [1$\bar{1}$5] directions in any of the three samples, and the
diffraction pattern about the (004) peaks is found to be invariant
with rotation about the growth axis [001], implying no tilt between
the substrates and the epilayers. \par

The samples are mounted in the variable temperature insert of a
magneto-optical cryostat. The electron spin dynamics are probed
with time-resolved Kerr rotation (KR)
spectroscopy$^{\cite{crooker1},\cite{crooker2}}$ in the Voigt
geometry using pulses from a frequency-doubled mode-locked
Ti:sapphire laser with a pulse duration $\sim$150 fs and a
repetition rate of 76 MHz. A circularly-polarized pump pulse
normal to the sample surface injects electrons with spins
polarized along the beam propagation direction \textit{z}.  After
a time delay \textit{$\Delta$t}, the Kerr rotation angle of a
linearly polarized probe pulse measures the projection of the
electron spin magnetization along \textit{z} as it precesses about
the transverse magnetic field \textit{B}$_{ext}$ applied along
\textit{y}. The pump and probe energies are typically tuned
between the range 2.70 - 2.80 eV, to maximize the KR effect, with
powers of 1 mW and 100 $\mu$W, respectively.  The beams are
focused to a spot size of $\sim$20 $\mu$m, and the circular
polarization of the pump beam is modulated with a photoelastic
modulator at 50 kHz for lock-in detection. Time-resolved
measurements$^{\cite{malajovich1}}$ have shown that the electron
g-factor for ZnSe in these samples is 1.1.\par

In order to characterize the internal magnetic field
\textit{B}$_{int}$, we measure KR as a function of the applied
magnetic field \textit{B}$_{ext}$ at a fixed pump-probe delay time
of \textit{$\Delta$t} = 13 ns while varying the applied electric
field \textit{E} and pump-probe separation \textit{d}. Fig. 2a
shows the data from the 100 nm sample (with the largest strain) at
\textit{T} = 125 K when the pump and probe are spatially
overlapped (\textit{d} = 0). The signal is an oscillatory function
of the total magnetic field $\vec{\textit{B}} =
\vec{\textit{B}_{ext}} + \vec{\textit{B}_{int}}$ and is given by
\textit{A}$\cos$(\textit{g}$\mu_{B}$\textit{B}\textit{$\Delta$t}/$\hbar$)$^{\cite{kato1},\cite{sih}}$
where \textit{A} is the amplitude, \textit{g} the effective
g-factor of the sample, $\mu_{B}$ the Bohr magneton and \textit{B}
the magnitude of the total magnetic field experienced by the
electrons. For KR along [110] in the absence of an applied
electric field (top trace), the average \textit{k}=0 and therefore
\textit{B}$_{int}$ = 0 and the oscillations are centered at
\textit{B} = \textit{B}$_{ext}$ = 0. For \textit{E} parallel to
\textit{B}$_{ext}$, the center peak is suppressed (center trace).
In this geometry, \textit{B}$_{int}$ is perpendicular to
\textit{B}$_{ext}$,$^{\cite{kato1},\cite{sih}}$ resulting in a
total field magnitude of \textit{B} =
$\sqrt{B_{ext}^{2}+B_{int}^{2}}$
 which is always greater than
zero for a non-zero \textit{B}$_{int}$. For larger applied
voltage, \textit{B}$_{int}$ increases, resulting in greater
suppression of the central peak. When the sample is rotated so
that \textit{E} is perpendicular to \textit{B}$_{ext}$ (bottom
trace), both the external and internal magnetic fields are along
\textit{y} and add directly so that \textit{B} =
\textit{B}$_{ext}$ + \textit{B}$_{int}$ and the oscillatory signal
is centered at -\textit{B}$_{int}$. It is noted here that all of
these effects are observed in the channel along [110] direction
only. For the channel along [1$\bar{1}$0], surprisingly no
internal fields in any geometry are observed in any of the
samples.

In both cases above, with non-zero electric field the amplitude of
the signal decreases with increasing voltage, which is further
investigated by spatially separating the pump and the probe by a
distance \textit{d} along the direction of \textit{E} (Fig. 2b).
Due to the laser profile of the pump beam, the optically injected
spins have a Gaussian spatial profile which is centered at
\textit{d} = 0 $\mu$m when \textit{E} = 0. An applied voltage
(\textit{E} $\neq$ 0) imparts a non-zero average momentum
\textit{k} to the injected spin packet, causing it to drift with
an average velocity \textit{v$_{s}$}.  We fit the amplitude
\textit{A} as a function of \textit{d} to determine the center
position \textit{d$_{c}$} = 6 $\mu$m of the spin packet with
applied voltage of 20 V after 13 ns, giving \textit{v$_{s}$} =
0.46 $\mu$m/ns.  Fig 2c shows the spatial variation of
\textit{B}$_{int}$ throughout the spin packet for \textit{E}
parallel to \textit{B}$_{ext}$. Spins at the leading edge of the
packet experience a larger \textit{B}$_{int}$ than the trailing
edge.  This variation is due to the spread in the drift velocities
of the spin packet arising from spin diffusion. The reported value
of \textit{B}$_{int}$ for each voltage are obtained from a linear
fit at \textit{d} = \textit{d$_{c}$}.\par

Fig. 3a shows both data and fits for \textit{A} (top) and
\textit{B}$_{int}$ (bottom) of the spin packet as measured by KR
at different applied voltages in the 100 nm sample at \textit{T} =
50 K. \textit{B}$_{int}$ increases linearly with applied voltage
(inset, top) confirming the \textit{k}-linear dependence. It also
increases with \textit{T}, almost doubling as \textit{T} changes
from 25 to 150 K (inset, bottom). It has been
observed$^{\cite{thomas}}$ that the in-plane biaxial strain in a
100 nm ZnSe/GaAs epilayer increases in magnitude with increasing
\textit{T} due to the thermal changes in the lattice and elastic
constants of both epilayer and substrate, which may explain the
changes observed here. However, without more systematic data on
the alteration of \textit{B}$_{int}$ with strain, this remains
speculative. While spin precession in these samples persist to
higher temperatures, at \textit{T} $>$ 150 K the spin lifetime is
not long enough to quantify the internal field.\par

Fig. 3b shows the variation of the spin-splitting energy $\Delta$ =
\textit{g}$\mu_{B}$\textit{B$_{int}$} with the spin drift velocity
\textit{v$_{s}$} for all samples at \textit{T} = 50 K. From linear
fits to the data, we use the slope $\beta$ =
$\Delta$/\textit{v$_{s}$} to characterize the samples and find
$\beta$ = 101, 86 and 103 neV ns/$\mu$m for the 300, 150 and 100 nm
samples, respectively. Previous measurements$^{\cite{sih}}$ in GaAs
samples with strain in a similar region of 2-3$\times10^{-3}$ have
shown $\beta$ values ranging between 75-125 neV ns/ $\mu$m.\par

The contact resistances in the samples are non-negligible and the
effective electric field across each channel cannot be simply
calculated from the applied voltage. Resistivity obtained from
transport measurements is used to calculate the electric field in
the two thicker samples, and in Fig. 3c we compare
\textit{B}$_{int}$ between the 300 and 150 nm samples as a
function of \textit{E}. Surprisingly, we observe the thicker
sample, with less strain, having a larger internal magnetic field
for the same electric field. Unfortunately, for the 100 nm sample
the contact resistance is too large to allow any reliable estimate
of the electric field\par

In conclusion, we have quantitatively measured the
\textit{k}-linear internal magnetic field as a function of voltage
in three samples. At this point we do not have a complete
understanding of the correlation between the internal magnetic
field and strain in ZnSe since comparison between the 300 and 150
nm samples show an increased spin splitting with decreasing
strain. This effect is not systematic, as we do not observe any
internal fields in the largely strain relaxed 1.5 $\mu$m thick
sample. We also observe \textit{B}$_{int}$ only along the [110]
direction in the samples while the strain is isotropic, which,
though not well understood, may be exploited to design spin
manipulation using topological effects$^{\cite{kato2}}$ .  \par

We thank NSF and ONR for financial support and H. Knotz and Y. Li
for assistance with the XRD measurements. N.P.S. acknowledges the
support of the Fannie and John Hertz Foundation.

\newpage

\begin{figure}\includegraphics{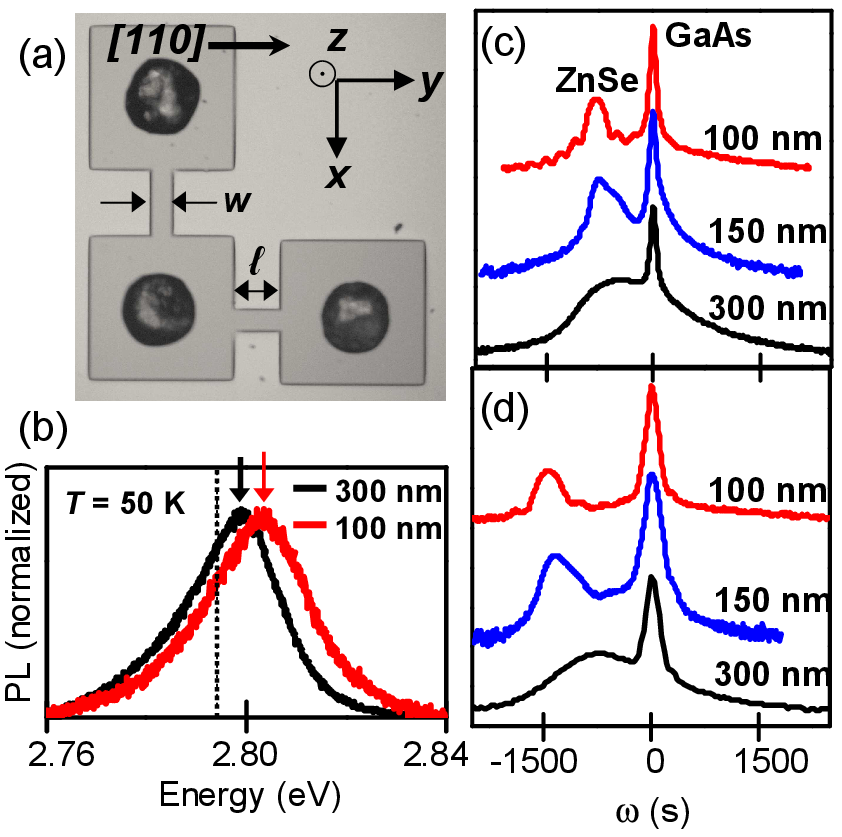}\caption{\label{fig1}
(a) Optical image of processed sample showing the crystal
orientation of the epilayer (\textit{l} = 235 $\mu$m, \textit{w} =
100 $\mu$m). The coordinate axes define the measurement geometry.
(b) PL for 300 and 100 nm samples showing the strain-induced shift.
Arrows indicate peak emission energies. Dashed line indicates the
bulk, unstrained energy gap. XRD spectra for the (c) (004) and (d)
(115) reflections. The appearance of interference fringes in the
(004) reflection in the 100 nm sample indicate phase coherent growth
and may be used to further verify the epilayer thickness.\\
}\end{figure}

\newpage

\begin{figure}\includegraphics{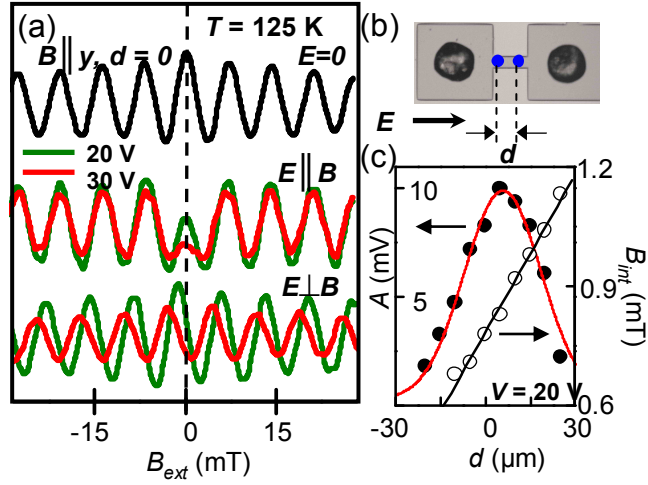}\caption{\label{fig2}
Data from 100 nm sample at \textit{T} = 125 K. (a) Kerr rotation
with pump-probe overlapped (\textit{d}=0) at \textit{$\Delta$t}  =
13 ns with (top) \textit{E} = 0, (center) \textit{E} $\parallel$
\textit{B}$_{ext}$ and (bottom) \textit{E} $\perp$
\textit{B}$_{ext}$. (b) Schematic of measurement geometry for
spatially resolved scans. (c) \textit{A} and \textit{B}$_{int}$,
obtained from fits described in the text, as a function of
pump-probe separation \textit{d} with \textit{E}$\parallel$
\textit{B}$_{ext}$.\\ }\end{figure}

\newpage

\begin{figure}\includegraphics{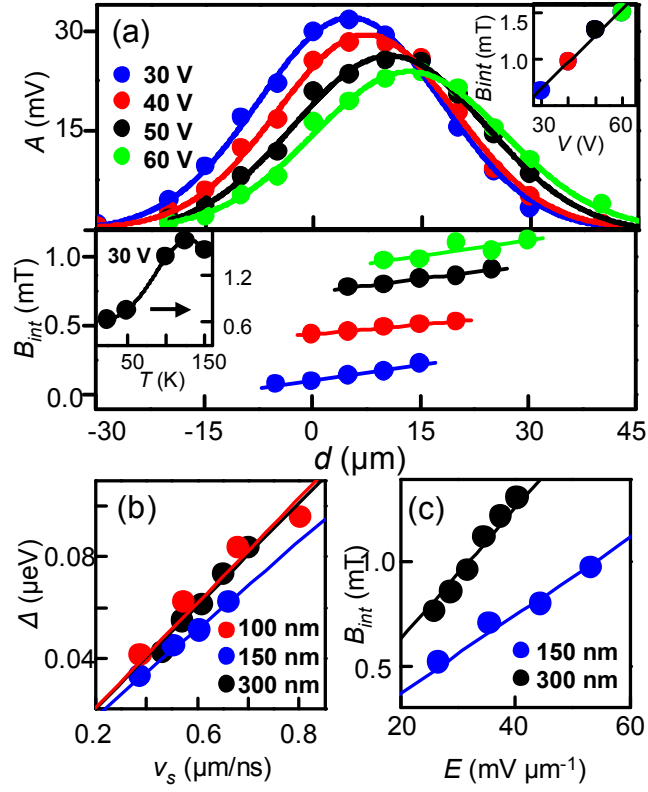}\caption{\label{fig3}
(a) \textit{A} and \textit{B}$_{int}$ (top and bottom) as a function
of pump-probe separation \textit{d} with
\textit{E}$\parallel$\textit{B}$_{ext}$, for a series of applied
voltages \textit{V} at \textit{T} = 50 K in 100 nm sample. (Inset,
top) \textit{B}$_{int}$ varies linearly with applied voltage.
(Inset, bottom) \textit{B}$_{int}$ as a function of temperature
\textit{T} for \textit{V} = 20 V. Comparison of (b) electric field
induced spin-splitting $\Delta$ at 50 K as a function of the spin
drift velocity \textit{v$_{s}$} for all samples and (c)
\textit{B}$_{int}$ for 300 and 150 nm samples varying with
\textit{E} at 50 K. }\end{figure}

\end{document}